\documentclass[12pt]{iopart}
\usepackage{hyperref}
\usepackage{graphicx}
\usepackage{epsfig}
\usepackage[utf8]{inputenc}

\newcommand{\ket}[1]{\mathop{\left| #1 \right\rangle}\nolimits}

\newenvironment{absolutelynopagebreak}
{\par\nobreak\vfil\penalty0\vfilneg
	\vtop\bgroup}
{\par\xdef\tpd{\the\prevdepth}\egroup
	\prevdepth=\tpd}

\begin{document}
\title{The Standard Quantum Limit \\ of Coherent Beam Combining}

\author{C~R~M\"{u}ller$^{1,2}$, F Sedlmeir$^{1,2}$, V~O~Martynov$^{3}$, Ch~Marquardt$^{1,2}$, and G~Leuchs$^{1,2,3}$
}
\address{	
$^1$ Max Planck Institute for the Science of Light, Erlangen, Germany.\\
$^2$ Department of Physics, Friedrich-Alexander-University Erlangen-Nuremberg (FAU), Germany. \\
$^3$ Institute for Applied Physics, Russian Academy of Sciences, Nizhny Novgorod, Russia. \\ 
}
\eads{\mailto{christian.mueller@mpl.mpg.de}
}

\date{\today}

\begin{abstract}
Coherent beam combining refers to the process of 
generating a bright output beam by merging independent input beams 
with locked relative phases.
We report the first quantum mechanical noise limit calculations 
for coherent beam combining and compare our results to 
quantum-limited amplification.
Our coherent beam combining scheme is based on an optical Fourier transformation 
which renders the scheme compatible with integrated optics. 
The scheme can be layed out for an arbitrary number of input beams 
and approaches the shot noise limit for a large number of inputs.
\end{abstract}


\maketitle

\section*{Introduction}
High power lasers are indispensable for various applications 
in fundamental research and industrial manufacturing. 
This pushed the development of high power fibre lasers.
Yet, since 2011, when thermal mode instabilities (TMI) were 
first observed \cite{Eidam11}, power scaling of fibre 
lasers came to a hold. 
TMI cause high power fibre lasers to become unstable above 
a certain threshold power and lead to chaotic fluctuations 
of their modal profile. Coherent beam combining (CBC) promises
power scaling beyond the TMI threshold. 
Multiple beams - typically split from a 
common seed laser - are individually amplified by identical 
fibre amplifiers operating below the TMI threshold and are 
then recombined, e.g., via constructive interference 
in a cascade of beam splitters, to generate the high power 
output beam \cite{Gaida15}. 

Apart from scaling the output power by combining the 
noisy outputs of amplifiers, 
CBC can also be applied to input beams with 
a quantum-limited noise profile. 
As we will see, the favourable noise 
scaling of CBC allows to extend the power range of 
(nearly) quantum noise-limited beams. 
Currently the power range is limited to the domain 
of merely a few Watts, which limits, e.g., the precision of 
quantum-limited metrology and the efficiency of coherent 
parametric processes.

Here we aim to elucidate the quantum limits of 
coherent beam combining paying particular attention to 
the scaling of the output noise as a function of the 
accuracy of phase locking and the number of combined beams.
We derive the quantum mechanical limits of coherent 
beam combining and report on the noise scaling when 
combining either quantum noise limited or 
noisy beams, such as those emerging from a linear amplifier. 

The paper is organized as follows.
In Sec.\ref{QLim_CBC_CS} the coherent 
combination of coherent states is introduced and the impact of 
a finite phase locking accuracy onto the output quadratures 
of the combined state are derived.
In Sec.\ref{Sec_Fourier_CBC} the standard quantum limit for the 
variance of the phase locking scheme is derived on the 
basis of an optical circuit realizing the discrete 
Fourier transformation and its inverse. 
Moreover, the quadrature variances of the combined state are derived with 
respect to the quantum limited phase locking accuracy.

In Sec.\ref{CompToLinAmp}, we compare the output noise in CBC to that 
of a quantum limited linear amplifier. 
In the Appendix Sec.\ref{SQL_RelPhase} we motivate the standard quantum limit for measuring the relative phase fluctuations between two states. 
Sec.\ref{LinAmpAnalysis} elucidates the quantum mechanical origin 
of the noise penalty in quantum-limited amplifiers and finally 
in Sec.\ref{MultiAmp} it is demonstrated that the 3\,dB limit 
does not apply to optical signals with fluctuations dominated by 
classical excess noise rather than by the quantum uncertainty.


\section{Quantum-limited coherent combination}
\label{QLimCBC_CS}

In coherent beam combining the optical power from several 
distinct beams is interferometrically merged into a single optical mode. 
Ideally, the output power is the sum of the input powers 
and thus an integer multiple of the single beam power. 
In this respect CBC can be seen as an amplifier 
with discrete steps in the gain. 
We emphasize, however, that in contrast to linear amplifiers, 
CBC requires multiple coherent input signals and \textit{a priori} 
knowledge on the signals amplitude as a resource. 

The potential of CBC schemes lies on the one hand in 
their capability to reach optical power levels beyond 
the limits of conventional amplifiers \cite{Eidam11}. 
On the other hand - and this is the property we want to highlight - 
CBC allows to prepare quantum noise limited outputs asymptotically.
In the following sections we will derive the standard quantum limit for 
the output noise in CBC.


\subsection{Formal description of noise in coherent beam combining}
\label{QLim_CBC_CS}

A CBC setup for the combination of $N$ beams can be 
envisioned as a symmetric beam splitter with $N$ 
input- and output ports each.

The interference between the inputs 
results in an output highly concentrated 
in a single port if the relative phases of the input states 
are adjusted to realize the discrete Fourier transform 
of the input modes as shown in \cite{BCIvL01}.

Let us consider the coherent combination of $N$ inputs. 
To achieve a stable output, the beams are mutually phase locked. 
The accuracy of the phase lock, however, is ultimately limited by 
quantum noise such that the optical phases, $\psi_{k}, k\in\left\{1,2,\cdots,  N \right\}$. 
fluctuate with a certain variance $\mathrm{Var}(\psi)$.
The input states are quantum mechanically described by their 
mode operators $\hat{a}_{k}$, which can be represented in a linearized 
form to single out the contributions of signal amplitude and quantum noise.
\begin{equation}
\hat{a}_{k} = \alpha \exp(i\,\psi_{k})+ \delta^{(2)}\hat{a}_{k}.
\label{eq:LinModeOp}
\end{equation}
$\alpha$ denotes the mean complex amplitude common to all input beams, 
$\exp(i\,\psi_{k})$ describes the phase fluctuations due to 
the (quantum-)limited phase-lock accuracy, and 
$\delta^{(2)}\hat{a}_{k} = \delta \hat{x}_{k} + i \delta \hat{p}_{k}$, 
denotes the quantum mechanical Heisenberg uncertainty 
along the conjugate quadratures 
$\hat{x} = \left(\hat{a}+\hat{a}^{\dagger}\right)/2$ and 
$\hat{p} = \left(\hat{a}-\hat{a}^{\dagger}\right)/(2i)$, 
$\left[\hat{a}, \hat{a}^{\dagger}\right]=1$.

The constructively interfering signal in the CBC output port 
is calculated by adding up the instantaneous amplitudes of the 
$N$ input beams.
\begin{equation}
\hat{a}_{out} = \sum_{k=1}^{N} \frac{\hat{a}_{k}}{\sqrt{N}}.
\label{eq:OutputModeOp}
\end{equation}
Note, that the mode operators are rescaled by $\sqrt{N}$ as 
the states are split symmetrically onto the $N$ output ports.

Following Eq.(\ref{eq:LinModeOp}) and Eq.(\ref{eq:OutputModeOp}), 
the output state is given by 
\begin{equation}
\hat{a}_{out} = \sum_{k=1}^{N} \left[\underbrace{\frac{\alpha\,\exp(i\,\psi_{k})}{\sqrt{N}}}_{\alpha_{\mathrm{out}}} +\underbrace{\frac{\delta^{(2)}\hat{a}_{k}}{\sqrt{N}}}_{\delta^{(2)}\hat{a}_{\mathrm{out}}}\right]. 
\label{eq:eq:LinearSummation}
\end{equation}

Let us discuss the constituents on the right hand side 
of Eq.(\ref{eq:eq:LinearSummation}) in more detail starting 
with the complex amplitude $\alpha_{out}$.
For phase-locked coherent combining we can assume the 
relative phase fluctuations to be small, $\mathrm{Var}(\psi) \ll 1$. 
This allows approximating the exponential phase fluctuation 
term via its Taylor series expansion to second order. 
The first order term represents 
a phase shift of the complex amplitude and the second order 
term describes a reduction of the output amplitude due to 
phase locking imperfections . 

\begin{equation}
\alpha_{\mathrm{out}} = \sum_{k=1}^{N} \frac{\alpha}{\sqrt{N}} \exp(i\,\psi_{k}) 
\approx \frac{\alpha}{\sqrt{N}} \sum_{k=1}^{N} \left(1 + i\,\psi_{k} - \,\frac{\psi_{k}^{2}}{2} \right).
\label{eq:AmplitudePart}
\end{equation}
Summing over the $N$ independent random phases $\psi_k$ 
with equal variance $\mathrm{Var}(\psi)$ yields a statistical 
distribution with variance $N\,\mathrm{Var}(\psi)$. 
The distribution of the squared random phases, $\psi^{2}_{k}$, 
is described by the gamma-distribution $\Gamma \left(k, \theta \right)$, 
with shape $k=N/2$ and scale $\theta = 2\,\mathrm{Var}(\psi)$. 
For reasonably large $N$, the gamma distribution can be 
approximated by its mean $k\theta=N\,\mathrm{Var}(\psi)$ 
and variance $k\theta^2 = 2N\mathrm{Var}^{2}(\psi)$. 
Within these approximations, the CBC output amplitude can be expressed as

\begin{eqnarray}
\alpha_{\mathrm{out}} &=& \frac{\alpha}{\sqrt{N}} \left(N + i\,\sqrt{N\,\mathrm{Var}(\psi)} - \frac{1}{2}\left[ N \mathrm{Var}(\psi) \pm \sqrt{2N} \mathrm{Var}(\psi) \right] \right) \nonumber \\
&=& \underbrace{\sqrt{N}\,\alpha \left(1 - \frac{\mathrm{Var}(\psi)}{2} \right)}_{\mathrm{mean\,value}} \pm \underbrace{\alpha\left(\frac{1}{\sqrt{2}}\mathrm{Var}(\psi) + i\sqrt{\mathrm{Var}(\psi)}\right)}_{\mathrm{fluctuations}}.
\label{eq:AmplitudePart2}
\end{eqnarray}

The second term in Eq.(\ref{eq:eq:LinearSummation}) concerns 
the evolution of the input states' quantum noise.  
The quantum fluctuations of coherent states can be modeled as 
independent Gaussian random variables, such that the variance 
of the combined fluctuation scales identically to the scaling of the 
phase fluctuations. 
Summing over the $N$ inputs and taking the beam splitting factor 
$1/\sqrt{N}$ into account yields another Gaussian 
random variable with the same variance as for the input beams. 
Hence, the quantum noise is preserved in the combining process.
\begin{equation}
\delta^{(2)}\hat{a}_{\mathrm{out}} = \frac{1}{\sqrt{N}} \sum_{k=1}^{N} \delta^{(2)}\hat{a}_{k} = \delta^{(2)}\hat{a}.
\label{eq:QuantumPart}
\end{equation}

Combining Eq.(\ref{eq:AmplitudePart2}) and 
Eq.(\ref{eq:QuantumPart}) yields the expression 
for the coherently combined output.
\begin{equation}
\hat{a}_{out} = \sqrt{N}\alpha \underbrace{\left(1 - \frac{\mathrm{Var}(\psi)}{2} \right)}_{\mathrm{amplitude\,reduction}} + \,\delta^{(2)}\hat{a} \pm \underbrace{\alpha\left(\frac{1}{\sqrt{2}}\mathrm{Var}(\psi) + i\sqrt{\mathrm{Var}(\psi)}\right)}_{\mathrm{noise\,from\,\,phase\,locking}}.
\label{eq:CombinedOutputStateCBC}
\end{equation}

\subsection{Fourier transform based coherent beam combining}
\label{Sec_Fourier_CBC}
In this section, we derive the standard quantum limit for 
the phase locking accuracy in combining $N$ coherent input states. 
The maximal accuracy with which two optical signals can be phase-locked 
is limited by the precision in estimating their relative phase.
The standard quantum limit for the minimal variance 
in estimating the relative phase of two coherent states depends 
solely on their mean photon number $n$.
\begin{equation}
\mathrm{Var}_{\mathrm{SQL}}(\psi) \geq \frac{1}{n}.
\label{eq:PhaseQLimCohState}
\end{equation}

\begin{figure}%
\centering
\includegraphics[width=.6 \columnwidth]{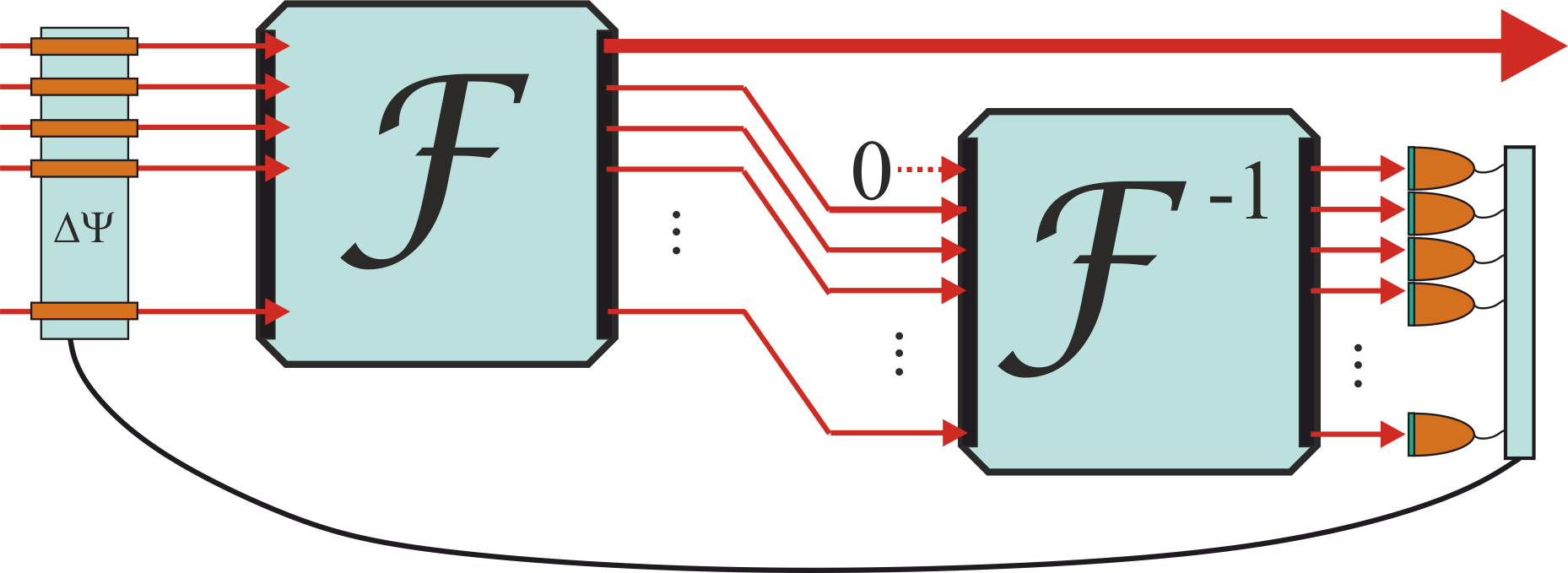}%
\caption{Fourier-transform based CBC scheme. See main text for details.}%
\label{fig:FourierScheme}%
\end{figure}

To extend this result to the combination of $N$ 
coherent states, we consider a CBC scheme based on the 
discrete Fourier transform. 
A sketch of this scheme is shown in Fig.\ref{fig:FourierScheme}.
The discrete Fourier transform of $N$ coherent input fields 
$\alpha_j = |\alpha|\exp(i \psi_j), \, j \in \{0, 1,..., N-1\}$ 
yields the outputs
\begin{equation}
\beta_{k} = \frac{1}{\sqrt{N}}\sum_{j=0}^{N-1} \alpha_{j} \, e^{-2\pi i \frac{j\,k}{N}}. 
\label{eq:DFT}
\end{equation} 

The first output port ($k=0$) readily provides the coherently 
combined beam $\beta_{0} = \sqrt{N}\,\bar{\alpha}$, 
where $\bar{\alpha}= \frac{1}{N}\sum_{j=0}^{N-1} \alpha_{j}$ 
is the average input amplitude. 
The relative optical phases of the input states are 
encoded as Fourier amplitudes in the beams exiting 
through the output ports with $1\leq k \leq N-1$. 
In the next step, we apply the inverse discrete Fourier 
transform to the corresponding outputs, where the extracted CBC port 
($k=0$) is left in the vacuum state. 
In essence, this scheme subtracts the instantaneous average 
value of the input beams $\bar{\alpha}$ from the 
individual inputs. 
The resulting outputs, hence provide the error signals 
$\Delta_j = \alpha_{j}-\bar{\alpha},\,j \in \{0,1,.., N-1\}$.
For small relative phases the error signals $\Delta_j$ are 
to a good approximation purely imaginary,
 $\alpha_{j}\approx i|\alpha_j|$, such that the phase difference 
between the individual states and the average state $\bar{\alpha}$ 
is directly encoded in the amplitude of the error signals. 
It is therefore not required to simultaneously measure both the 
$X$ and the $P$ quadrature and single photon detection is sufficient 
to extract the required information.
The sensitivity of detecting a phase difference at the output ports 
of the inverse Fourier transformation is ultimately limited by 
quantum mechanics. 

Due to the stochastic nature of photon measurements in each channels it is not possible to perform perfect phase correction based on a single detector click. Properly designed feedback scheme will decrease the input phase mismatch down to a certain limit after some number of clicks.
In this paper, we assume an "ideal" feedback that improves phase difference after each click of any detector. For such a scheme we argue that the limit corresponds to an error signal with mean photon number of one when summed over all output ports.

\begin{eqnarray}
n_{\mathrm{err}} &=& \sum_{j=0}^{N-1} |\Delta_j|^2
=\sum_{j=0}^{N-1} |\alpha \exp(i \psi_{j}) - \bar{\alpha} |^2 \nonumber \\
&=&\sum_{j=0}^{N-1} |\alpha \exp(i \psi_{j}) - \frac{1}{N}\sum_{k=0}^{N-1} \alpha \exp(i \psi_{k}) |^2 \nonumber \\
&\approx& n\cdot \sum_{j=0}^{N-1} |(1 + i \psi_{j}) - \frac{1}{N} \sum_{k=0}^{N-1} (1 + i \psi_{k}) |^2 \nonumber \\
&=& n\cdot \sum_{j=0}^{N-1} (\psi_{j} - \frac{1}{N} \sum_{k=0}^{N-1} \psi_{k})^2 \nonumber \\
&=& n\cdot \sum_{j=0}^{N-1} (\psi_{j} - \langle \psi \rangle)^2,
\label{eq:FourierLimit}
\end{eqnarray}
By using the Taylor approximation we have arrived at an expression 
containing the sum over the squared deviations of the individual phases from the instantaneous average phase $\langle \psi \rangle$.
This last line in Eq.\ref{eq:FourierLimit} is $n (N-1)$ times the unbiased sample variance of the relative input phases.
\begin{equation}
n \sum_{j=0}^{N-1} (\psi_{j} - \langle \psi \rangle )^2 = n\,(N-1)\,\mathrm{Var}(\psi).
\label{eq:FourierLimit2}
\end{equation}

The requirement of detecting at least one photon on average 
finally yields the standard quantum limit for the input 
states' minimal phase uncertainty.
\begin{equation}
n\,(N-1)\,\mathrm{Var}_{\mathrm{SQL}}(\psi) = 1 \Rightarrow \mathrm{Var}_{\mathrm{SQL}}(\psi) = \frac{1}{(N-1)\,n}
\label{eq:FourierLimit3}
\end{equation}

Imperfections in the phase-locking accuracy can be captured 
by introducing an accuracy factor $\xi$, 
where the minimal value $\xi = 1$ corresponds to the quantum limit: 
$\mathrm{Var}(\psi) = \xi\,\mathrm{Var}_{\mathrm{SQL}}(\psi) 
= \frac{\xi}{(N-1)\,n}$.

The output noise can be divided into 
independent fluctuations along the in-phase quadrature $\Delta x = \alpha\,\Delta^2\psi'/\sqrt{2}$, 
corresponding to the amplitude direction, and the out-of-phase quadrature $\Delta p = i\alpha\,\Delta\psi$.
Using the expression for the quantum limited phase variance 
and considering the coherent state quadrature variance 
$\mathrm{Var}_{coh} = 1/4$ (see Eq.(\ref{eq:Vcoh}) in the Appendix), 
the resulting noise variance along the quadratures yields
\begin{eqnarray}
\mathrm{Var}_{out}^{(CBC)}(x)  &=& \mathrm{Var}_{coh} + \frac{n}{2}\,\left(\xi\,\mathrm{Var}_{\mathrm{SQL}}(\psi)\right)^2  \nonumber \\
&=& \left( 1+ \frac{2 \xi^2}{(N-1)^2\,n}\right) \mathrm{Var}_{coh}, \nonumber \\
\mathrm{Var}_{out}^{(CBC)}(p) &=& \mathrm{Var}_{coh} + n\,\xi\,\mathrm{Var}_{\mathrm{SQL}}(\psi) \nonumber \\
&=& \left( 1+ \frac{4 \xi }{N-1}\right) \mathrm{Var}_{coh},
\label{eq:VarAmpOut}
\end{eqnarray}
where $n=|\alpha|^2$. 
The standard quantum-limit for the output noise in 
coherent beam combining corresponds to $\xi = 1$. 
The magnitudes of the excess noise variance $\mathrm{Var}_{x,p}^{\mathrm{excess}}$, 
i.e. excluding the shot noise, 
are not symmetric along the quadratures but are dominated by phase noise
$\mathrm{Var}_p^{\mathrm{excess}} = 2 n (N-1) \mathrm{Var}_x^{\mathrm{excess}}$ . 
For a large number of input states $N\gg1$ both variances approach the shot noise level.


\section{Comparison between the quantum limits in linear amplification 
and Coherent Beam Combining} \label{CompToLinAmp}

An amplifier can be defined as a device that 
takes an input signal, increases its power and outputs a 
signal that is a rescaled version of the input. 
The amplification gain $G$ is defined as the 
power ratio between the output and the input signal. 
In the context of linear phase-insensitive amplifiers, 
one is often confronted with the famous 3\,dB limit \cite{Caves82}. 
In vague terms, the 3\,dB limit states, that the 
amplification of a signal with high gain $G$ increases the 
field quadrature variances by a factor $2G$.
The exact equation for the output quadrature variance of 
a quantum limited linear amplifier is
\begin{equation}
\mathrm{Var}_{amp} = (2G-1) \mathrm{Var}_{coh}, 
\label{eq:NoiseLimitLinAmp}
\end{equation}
The additional excess noise impedes many quantum 
optical applications that rely on quantum-limited noise 
properties and has profound implications on the maximal 
distance between repeater stations of optical communication 
channels \cite{essiambre2012capacity}.
The notion of noiseless amplification bears some ambiguity. 
Depending on the context, it may refer to an amplifier 
preserving the quantum state's quadrature variances 
$\mathrm{Var}_{amp} = \mathrm{Var}_{in}$.
Such an amplifier would apply a coherent displacement 
along the signal's exact complex amplitude, which can only 
be realized if the phase of the input state is known \textit{a priori} 
and is unphysical otherwise. 
Here we refer to a noiseless amplifier as one for which 
the signal's quadrature variance scales equally to its 
mean photon number $\mathrm{Var}_{amp} = G\,\mathrm{Var}_{in}$, 
such that the noise figure does not change. 
The noise figure is the ratio of the signal-to-noise ratios SNR$_{in}$ 
and SNR$_{out}$.
This amplifier model could be realized by a phase-sensitive amplifier, 
and is also approximately valid for the phase-insensitive amplification of 
input states with classical noise significantly exceeding 
the quantum noise limit (see Appendix Sec.\ref{MultiAmp} for details).

\begin{figure}[htb]%
\centering
\includegraphics[width=.5\columnwidth]{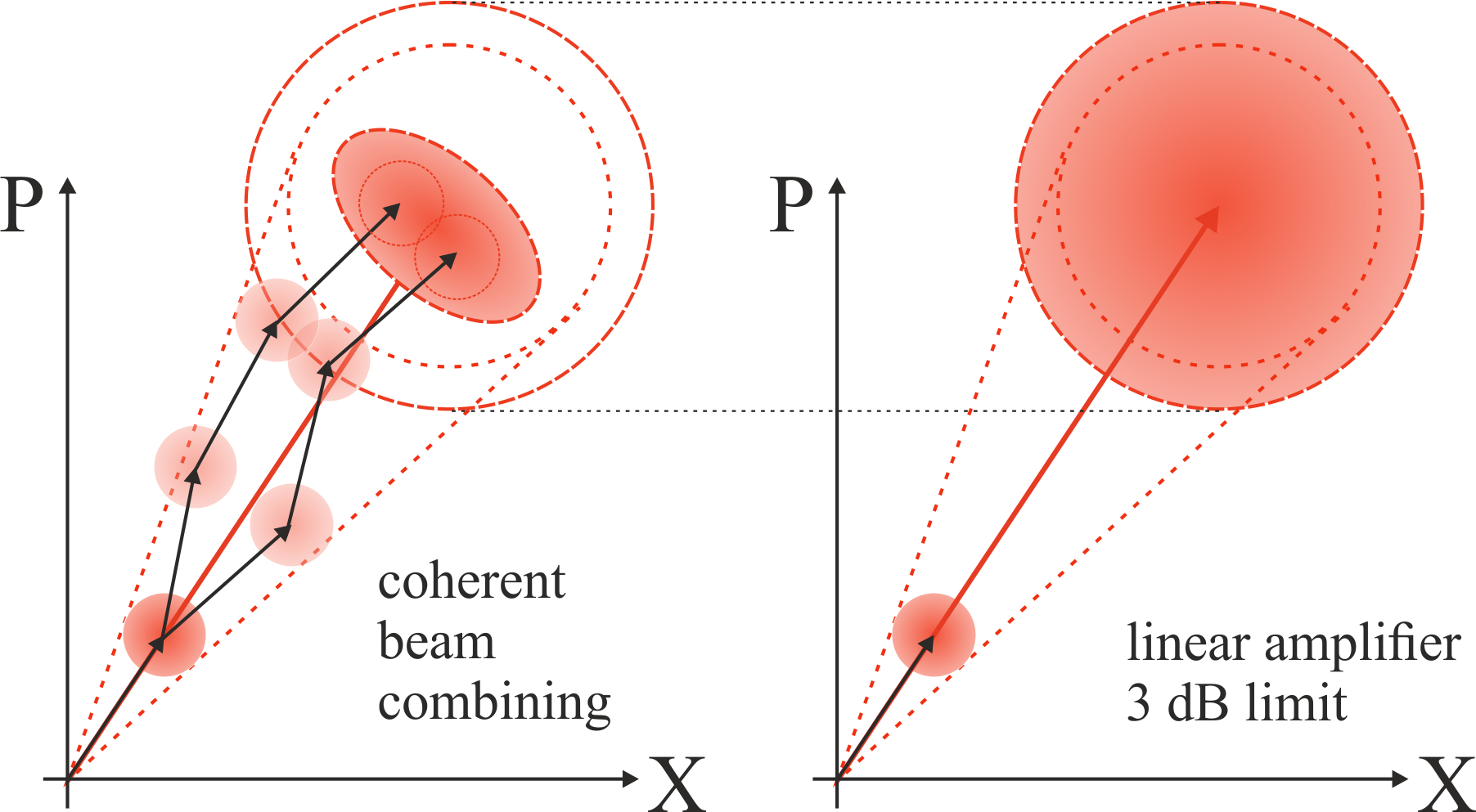}%
\caption{Phase space sketch comparing the excess noise scaling 
of coherent beam combining (left) and of linear amplification (right). 
In the linear amplifier, the quadrature variance is proportional to the gain factor $G$. 
In contrast, the quadrature variances approach asymptotic values in case of CBC, as
the independently added noise contributions progressively average out.}%
\label{fig:sketch_amp_vs_CBC}%
\end{figure}

Let us finally compare the quantum limits for the 
noise scaling of CBC and linear amplification in detail.
The minimal noise variance of a quantum limited 
linear amplifier is phase symmetric and proportional 
to the gain factor $G$ as outlined in 
Eq.(\ref{eq:NoiseLimitLinAmp}). 
The noise scaling in CBC is fundamentally different. 
In adding up multiple input beams, the 
phase fluctuations from the imperfect phase locking 
are progressively averaged out as sketched in 
Fig.\ref{fig:sketch_amp_vs_CBC}. 

In stark contrast to the
linear amplifier, the output noise is not 
proportional to the gain factor $G=N$, but scales 
inversely with the number of combined states
(see Eq.(\ref{eq:VarAmpOut})).
This statistical noise suppression results in a 
significantly reduced noise footprint compared to 
a linear amplifier. 
Let us analyze how large a relative phase fluctuation 
$\Delta^{2} \psi$ can be tolerated in CBC before the 
output variance surpasses that of a quantum-limited 
amplifier $\mathrm{Var}^{(amp)}_{out}(N)$.
As the output noise is typically dominated by the 
fluctuations along the out-of-phase quadrature $p$, 
we choose the corresponding variance as the 
conservative reference for the comparison. 

\begin{absolutelynopagebreak}
\begin{eqnarray}
\mathrm{Var}_{SQL}^{(CBC)}(p) = \left(1 + \frac{4\xi}{N-1}\,\xi \right)\mathrm{Var}_{coh} \stackrel{!}{=}  (2N-1)\,\mathrm{Var}_{coh}= \mathrm{Var}^{(amp)}_{out} \nonumber \\
\Rightarrow \xi = \frac{1}{2}\,(N-1)^2.
\label{eq:SingleN_threshold}
\end{eqnarray}
\end{absolutelynopagebreak}
The accepatable phase variance scales quadratically 
in the number of combined states.
Hence, CBC can outperform the output noise 
variances of a linear amplifier irrespective of the 
phase locking accuracy, if only the number of combined 
signals is sufficiently large. 


\section*{Conclusion}
Coherent beam combining is an enabling technology for the 
generation of brightest beams with excellent beam quality. 
Our analysis reveals that CBC also exhibits 
intriguing properties regarding the quantum limited 
noise scaling in the combination of both pure and noisy input beams.

\section*{Acknowledgments} 
This work is funded by the the Fraunhofer and
Max-Planck Cooperation Program (PowerQuant) and by the Ministry of Science and Higher Education of the Russian Federation (Contract No. 14.W03.31.0032).
The authors thank Cesar Jauregui, Jens Limpert, 
Nicoletta Haarlammert and Thomas Schreiber for fruitful discussions.

\section*{References}
\bibliography{References}
\bibliographystyle{unsrt}

\section{Appendix}


\subsection{Standard quantum limit for relative phase measurements}
\label{SQL_RelPhase}

CBC is based on constructive interference, such that
the noise profile of the combined beam depends crucially 
on the relative optical phases of the input beams. 
The precision of phase locking is fundamentally 
limited by the Heisenberg uncertainty principle. 
\begin{equation}
\Delta x \Delta p \geq \frac{1}{4}.
\label{eq:HeisenbergXP}
\end{equation}
For coherent states the quantum uncertainty is 
distributed symmetrically onto the quadrature axes.
\begin{equation}
\Delta^2 x = \Delta^2 p = \mathrm{Var}_{coh} = \frac{1}{4}.
\label{eq:Vcoh}
\end{equation}
An overview on the quantum noise limits in optical 
interferometry can be found in  \cite{Berry09, Demkowicz15, Dowling15}. 
The ultimate phase estimation precision compatible with 
the physical laws of quantum mechanics is given by the 
Heisenberg limit is $\Delta \psi_{HL} = 1/n$ \cite{Holland93}
which, however, can only be achieved using exotic quantum states 
such as N00N states. 
The optimal precision for estimating the relative phase between 
coherent states is given by the standard quantum limit
$\Delta \psi_{SQL} = 1/\sqrt{n}$ \cite{Caves81}. 

\begin{figure}[htb]
\centering 
\includegraphics[width=.80\columnwidth]{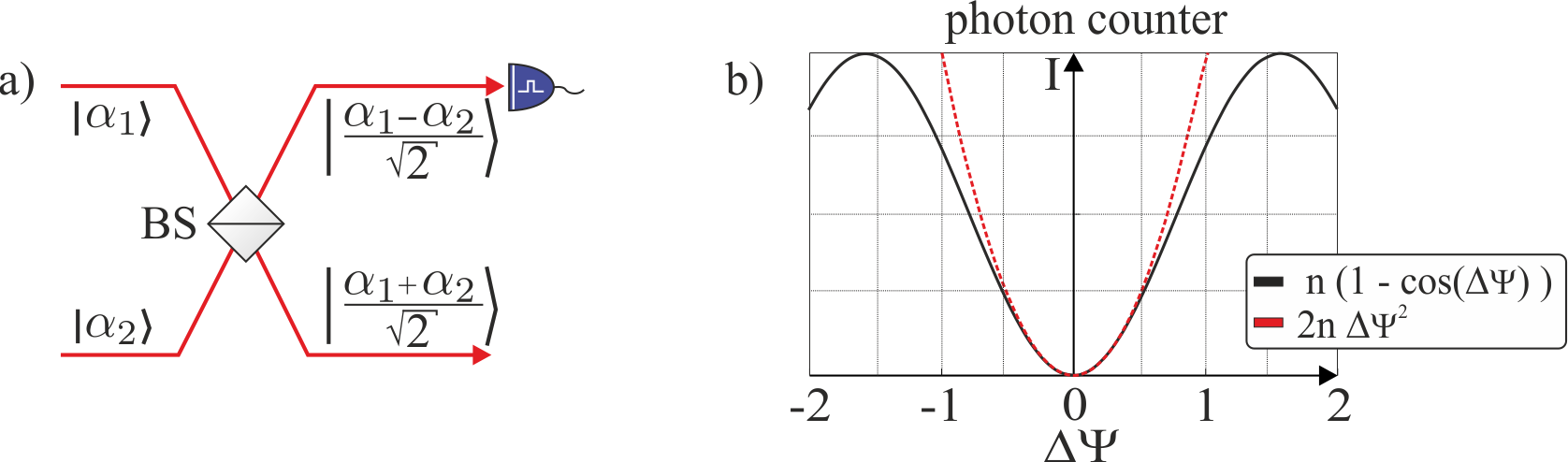}%
\caption{
a) Schematic depicting the coherent combination of two coherent states. 
b) Single photon detection in the minus port provides information about 
relative phase. For small phase fluctuations the curve can be approximated 
by a parabola which directly yields the accuracy at the SQL.
}%
\label{fig:interferometer_phase_sensitivity}%
\end{figure}

A CBC locking scheme satisfying the standard quantum limit for the 
relative phase accuracy can be realized using a single photon detector 
and is sketched in Fig.\ref{fig:interferometer_phase_sensitivity}\,a).
Two coherent beams $\ket{\alpha_{1}}$ and $\ket{\alpha_{2}}$ 
interfere on a symmetric beam splitter and are combined in the 
output port with constructive interference 
Information about the relative phase of the coherent states 
can be extracted from the optical power exiting the beam splitter 
through the destructive interference port.
Ideally, the amplitudes and the signal phases are 
equal such that the photon counter in this port detects 
the vacuum state. 
The error signal in the minus port is depicted in 
Fig.\ref{fig:interferometer_phase_sensitivity}\,b).
The minimal phase shift $\Delta\psi$ that can be detected 
between the input beams is determined by the requirement 
that at least a single photon is detected. 
\begin{eqnarray}
\langle \hat{n}_{-}(\Delta \psi) \rangle &=& |\alpha|^2\,\langle(1-\cos(\Delta\psi)\rangle) \approx n \langle(1 - (1-\frac{\psi^2}{2})\rangle) \stackrel{!}{=} 1 \nonumber  \\
\nopagebreak[4] \Rightarrow \Delta^2 \psi_{min} &=& \frac{2}{n},
\label{eq:APDmeanSignal}
\end{eqnarray}
where we used the Taylor approximation for the 
cosine function as we assume small $\psi$.
In this derivation one of the states is modeled 
as having a fixed phase and all phase fluctuations 
are attributed to the second state. 
This can straightforwardly be symmetrized by 
equally distributing the variance among the two inputs. 
This yields the standard quantum limit for the minimal 
detectable phase variance of each of the input states 
$\Delta \psi_{SQL} = \frac{1}{n}$.

\subsection{Noise scaling in linear amplifiers}
\label{LinAmpAnalysis}
In the canonical quantization of the electromagnetic 
field \cite{LeonhardtMeasuring}, the photon annihilation 
operator $\hat{a}$ is identified as the quantum 
counterpart to the classical amplitude variable $\alpha$.
This suggests the following operator description for the 
amplified signal $\hat{a}_{out} = g \, \hat{a}_{in}$.
Both, the input 
$\hat{a}_{in}$ and the output $\hat{a}_{out}$ 
refer to optical modes, which for a proper 
quantum transformation are connected by a 
unitary transformation. As the commutator 
bracket is invariant under unitary transformations 
a necessary condition for the amplifier output 
$\hat{a}_{out}$ is that the canonical 
commutation relation is preserved 
$\left[ \hat{a}_{out}, \hat{a}^{\dagger}_{out} \right]\stackrel{!}{=}1$. 
From the naive approach, however, we obtain
\begin{equation}
\left[ \hat{a}_{out}, \hat{a}_{out}^{\dagger}\right]= g \left[ \hat{a}_{in}, \hat{a}_{in}^{\dagger}\right] g = g^2 = G, 
\label{eq:CommutatorNaive}
\end{equation}
which violates the canonical commutation relation 
for any nontrivial gain $g\neq1$. 
Yet, the commutation relation can be preserved 
by introducing an additional \textit{noise term} $\hat{N}$
\begin{eqnarray}
\left[ \hat{a}_{out}, \hat{a}^{\dagger}_{out}  \right] &=& \left[ g \, \hat{a}_{in} + \hat{N}, g \,\hat{a}^{\dagger}_{in} + \hat{N}^{\dagger}  \right]  \nonumber \\
&=& g^2\,\underbrace{\left[ \hat{a}_{in}, \hat{a}^{\dagger}_{in}  \right]}_{=\,1} + \left[ \hat{N}, \hat{N}^{\dagger}  \right] + g\underbrace{\left[\hat{N}, \hat{a}^{\dagger}_{in} \right] }_{=\,0} + g\underbrace{\left[\hat{a}_{in}, \hat{N}^{\dagger}  \right] }_{=\,0}.
\label{eq:ampcomm}
\end{eqnarray}
The amplitude operator of the input signal $\hat{a}_{in}$ 
and the noise term $\hat{N}$ act on different subsystem - 
the optical field mode and the internal modes of 
the amplifier, respectively. 
As the amplifier should not be biased towards a 
certain input signal, the optical mode 
$\hat{a}_{in}$ and the internal modes of the 
amplifier $\hat{N}$ cannot be correlated and 
must commute. 
Consequently, in order to preserve the 
bosonic commutation relation for $\hat{a}_{out}$, 
the commutator bracket of the noise term must satisfy 
$\left[ \hat{N}, \hat{N}^{\dagger} \right]=1-g^2$, 
which allows to express the \textit{noise term} 
as an operator of the form
\begin{equation}
\hat{N} = \sqrt{g^2-1}\,\hat{b}^{\dagger} + N_{cl}.
\label{eq:NoiseTermDefinition}
\end{equation}
$\hat{b}$ denotes a bosonic ancilla mode with vanishing mean amplitude $\langle \hat{b} \rangle = 0$. Note, that the commutator equation could in principle also be satisfied if we set $\hat{N} = \sqrt{1-g^2}\,\hat{b}$. For an amplifier, however, we require $g^2 > 1$, such that $\left[\sqrt{1-g^2}\,\hat{b}, (\sqrt{1-g^2}\,\hat{b})^{\dagger} \right] = \underbrace{|\sqrt{1-g^2}|^2}_{>0} \stackrel{!}{=} \underbrace{1-g^2}_{<0}$ cannot be satisfied.
The additional c-number $N_{cl}$ can be introduced to account for any classical imperfections of the amplifier. 
For amplifiers working at the quantum limit $N_{cl} = 0$. \newline
The fundamental input-output relation for the quantum limited 
phase-insensitive linear amplifier follows as
\begin{equation}
\hat{a}_{out} = g\,\hat{a}_{in} + \sqrt{g^2-1}\,\hat{b}^{\dagger}.
\label{eq:amplifierINOUT}
\end{equation}
In order to quantify the performance of the amplifier, we calculate the output variance for a coherent input state $\ket{\alpha}$. 
Coherent states are symmetric minimum uncertainty states with quadrature uncertainties $\mathrm{Var}(x) = \mathrm{Var}(p) = 1/4$. 
In a phase-insensitive linear amplifier both quadrature variances are increased equally. 
Therefore, we can restrict the analysis to the variance of the $X$ quadrature $\mathrm{Var}(x) = \langle \hat{X}^2 \rangle - \langle \hat{X} \rangle ^2$. 
\begin{equation}
\mathrm{Var}(x_{out}) =  \langle \left( \frac{\hat{a}_{out} + \hat{a}_{out}^{\dagger}}{2} \right)^2 \rangle -  \langle \frac{\hat{a}_{out}+\hat{a}_{out}^{\dagger}}{2} \rangle^2
\end{equation}
Applying Eq.(\ref{eq:amplifierINOUT}), the output variance can be decomposed into three terms $\mathrm{Var}(x_{out}, g) = \Gamma_{sig}(g) + \Gamma_{noise}(g) + \Gamma_{cross}(g)$. 
The first term describes the rescaling of the input signal's variance. 
\begin{eqnarray}
\Gamma_{sig}(g) &=& \langle \left( g\, \frac{\hat{a}_{in} + \hat{a}_{in}^{\dagger}}{2} \right)^2 \rangle -  \langle g\,\frac{\hat{a}_{in}+\hat{a}_{in}^{\dagger}}{2} \rangle^2  \\ 
&=& g^2\,\langle \hat{X}_{in}^2 \rangle - g^2\,\langle \hat{X}_{in}\rangle^2 = g^2 \cdot \mathrm{Var}(x_{in})  \nonumber
\end{eqnarray}
Note, that this term is solely dependent on the signal mode $\hat{a}$ and describes the linear rescaling of both the signal power and the noise with equal gain $g^2$. 
Even though the modes of the amplifier $\hat{b}$ are not represented in this term, the variance is already increased. 
However, in the absence of the additional noise term, the SNR is still preserved. \newline 
The second term is responsible for the reduction of the SNR. 
Assuming that the internal mode of the amplifier is initially in the vacuum state and noting that the quadrature uncertainty of the vacuum state is identical to that of a generic coherent state we obtain
\begin{eqnarray}
\Gamma_{noise}(g) &=& \langle \left( \sqrt{g^2-1}\,\,\frac{\hat{b}^{\dagger}+\hat{b}}{2} \right)^2 \rangle - \underbrace{\langle \sqrt{g^2-1}\,\,\frac{\hat{b}^{\dagger}+\hat{b}}{2}  \rangle^2}_{=(g^2-1)\,\langle \hat{x}_b \rangle^2\,=\,0} \\
&=& (g^2-1)\,\,\mathrm{Var}(x_{in}) = (g^2-1)\,\,\mathrm{Var}(x_{vac}) \nonumber
\end{eqnarray}
The last term is a cross-term between the signal operator and the noise operator which vanishes as we require $\langle \hat{b} \rangle = 0 $.
\begin{equation}
\Gamma_{cross}(g) = \frac{g\,\sqrt{g^2-1}}{4}\,\langle \left(\hat{a}_{in}+\hat{a}_{in}^{\dagger}\right)\rangle \langle \underbrace{\left(\hat{b}^{\dagger}+\hat{b}\right)\rangle }_{= 2\langle \hat{x}_b\rangle = 0} = 0, 
\end{equation}
Note, that the expectation values for the signal and the 
noise term in the product could be evaluated independently 
as they are required to be uncorrelated. 
In sum we arrive at a lower limit for the output 
variance of the field quadratures
\begin{equation}
\mathrm{Var}(x_{out}) \geq  (2g^2 - 1)\,\mathrm{Var}(x_{in}).
\label{eq:QuadratureOutputNoise}
\end{equation}

Let us try to get some intuition on the factor $2g^2-1$. 
A simple, conventional amplifier can be composed of a quantum receiver performing a measurement on the coherent input state $\ket{\alpha}$ to derive an estimate $\tilde{\alpha}$ of the complex amplitude. 
Subsequently, a signal source prepares a rescaled coherent state $\ket{g\tilde{\alpha}}$ according to the obtained estimate. 
The origin of the excess noise in this example is founded in the impossibility to precisely determine the complex amplitude $\alpha$. 
The measurement can, for instance, be performed with a heterodyne detector or a double homodyne detector which both measure the conjugate field quadratures simultaneously.
Such measurements come at the price of adding at least one unit of vacuum noise to the measurement outcome \cite{Arthurs65, Stenholm92}. 
Consequently, the variance of the estimates' distribution 
is at least twice the quadrature variance of the coherent 
state $\mathrm{\tilde{V}} \geq  2\,\mathrm{Var}_{coh}$.
Applying the amplification factor to this distribution 
linearly increases its variance as 
$g^2\,\mathrm{\tilde{V}} \geq g^2\,2\,\mathrm{Var}_{coh}$. 
Preparing a coherent state at the phase space coordinate 
according to the amplified estimator finally adds another unit 
of shot noise $\mathrm{Var}_{coh}$ to the distribution, i.e. the 
Heisenberg uncertainty of the prepared quantum state itself. 
In sum this yields the output noise of Eq.(\ref{eq:QuadratureOutputNoise}). 

In contrast, the output noise of a quantum-limited amplifier 
is only $(2g^2-1)$-times as high as the coherent input noise. 
In the regime of low gain, a quantum-limited amplifier 
offers an appreciable performance gain. 
However, for high gain both the quantum-limited amplifier 
and the measure\&prepare amplifier asymptotically 
approach the $3\,$dB limit.

It is worth noting that the aforementioned 
noise limit only applies to deterministic and phase-insensitive 
amplifiers, which successfully amplify each signal state 
and which operate independent of the signal phase.
While these properties indeed apply to the notion of 
a conventional amplifier, alternative concepts such as 
phase-sensitive amplifiers and probabilistic amplifiers 
may offer advantageous noise properties (depending on the 
targeted application) and have also been demonstrated experimentally.
Phase-sensitive amplifiers \cite{Levenson93, McKinstrie04} 
exhibit a phase dependent gain profile. In its simplest
form a phase-sensitive amplifier provides (positive) 
gain only along one preferred quadrature, while the conjugate 
quadrature gets squeezed and is attenuated. 
Such an amplifier can be useful if, for example, the set 
of input states is aligned only along a single quadrature. 
An important example is the amplification of states from 
the binary phase-shift keyed (BPSK) alphabet \cite{Corcoran12}. 
Probabilistic amplification has been demonstrated to allow 
for noiseless amplification of coherent states in various 
experimental settings 
\cite{Usuga2010, Zavatta2010, Ferreyrol2010, Xiang2010}. 
All these schemes achieve noiseless amplification via 
a probabilistic nonlinear interaction and measurement that 
allows to herald successfully amplified signals. 
In probabilistic amplifiers the purity of the state 
is traded off against the success probability.


\subsection{Error signals in the Fourier Coherent Beam Combining Scheme} 
\label{FCBC_error}
In this section we provide a detailed calculation of the 
error signal amplitudes $\epsilon_{l}$ at the output of the inverse Fourier transformation. 
The separation of the coherently combined signal after the initial Fourier transformation 
as captured by the term $\left(1-\delta_{k,0} \right)$ at the end of the first line in Eq.\ref{eq:FCBC_error}

\begin{eqnarray}
\epsilon_{l} &=& \frac{1}{N} \sum_{k=0}^{N-1} \underbrace{\left(\sum_{n=0}^{N-1} \alpha_{n} \e^{-\frac{2\pi i}{N}\,k\,n}\right)}_{\mathcal{F}(\alpha)_{k}} \e^{\frac{2\pi i}{N}\,l\,k} \left(1-\delta_{k,0} \right) \nonumber \\
&=& \frac{1}{N} \sum_{n=0}^{N-1} \alpha_{n} \left[ \sum_{k=0}^{N-1} \e^{-\frac{2\pi i}{N}\,k\,(n-l)} -1\right] \nonumber \\
&=& \frac{1}{N} \sum_{n,k=0}^{N-1} \alpha_{n} \e^{-\frac{2\pi i}{N}\,k\,(n-l)} - \frac{1}{N} \sum_{n=0}^{N-1} \alpha_{n} \nonumber \\
&=& \frac{1}{N} \sum_{n=0}^{N-1} \alpha_{n} \delta_{n,l} - \bar{\alpha} \nonumber \\
&=& \alpha_{l} - \bar{\alpha} 
\label{eq:FCBC_error}
\end{eqnarray}

Hence the error signal amplitude at each output port $\epsilon_{l}$ 
is exactly the difference between the input amplitude $\alpha_{l}$ 
and the instantaneous mean amplitude $\bar{\alpha}$. 


\subsection{Quantum-limited linear amplification by multiple amplifier stages} 
\label{MultiAmp}
\begin{figure}[htb]
\centering
\includegraphics[width=.8\columnwidth]{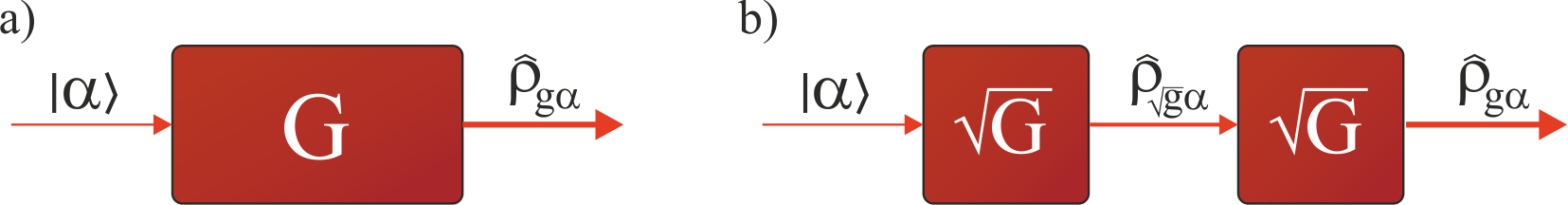}%
\caption{
Analog representations for the linear amplification of an input state. \newline
a) The input signal $\ket{\alpha}$ is amplified with (intensity) gain $G$ to 
a signal with mean amplitude $\alpha_{\mathrm{amp}} = g \alpha$. \newline
b) The input signal is amplified by two distinct amplifiers each with gain $\sqrt{G}$.}%
\label{fig:multiple_amplifier_stages}%
\end{figure}

It is a common misconception that high gain linear amplification is 
unconditionally linked to a 3\,dB reduction of the SNR. 
However, this noise penalty only applies to the 
amplification of pure signals (in the quantum mechanical sense). 
Let us shed some light on this by considering the 
amplification of a coherent state $\ket{\alpha}$ by either 
a single amplifier with intensity gain $G$ or alternatively by two 
consecutive amplifiers with gain $\sqrt{G}$ each 
(see Fig.\ref{fig:multiple_amplifier_stages}). 
Clearly, the mean amplitude of the amplified signals is 
identical in both cases. 
One could, however, assume that due to the 3\,dB noise 
penalty of linear amplifiers the latter scheme suffers 
from double the noise power as two individual amplification 
steps were invoked. 

For the single amplifier, the variance of the output state scales with the 
factor $(2G-1)$ which yields the famous 3,dB noise penalty for $G\gg 1$.
At this point, it is crucial to understand how the scaling factor comes about. 
In fact, the usual representation obfuscates the physics behind the amplification 
process, such that it is helpful to rewrite the factor as 
\begin{equation}
\mathrm{Var}_{\mathrm{amp1}} = (2G-1)\,\mathrm{Var}_{\mathrm{coh}} = \left( 1 + 2(G-1) \right)\,\mathrm{Var}_{\mathrm{coh}}
\label{eq:single_amp}
\end{equation}
In this form the scaling of the output state's variance is divided in two 
parts. 
The initial '1' represents the variance of the symmetric Heisenberg uncertainty 
relation for bosonic modes which remains unaltered by the amplification process. 
The second term $2(G-1)$ describes the additional variance due to the scaling 
of the input variance in the amplification process. 
The factor of 2 is due to the simultaneous amplification of the two conjugate 
quadratures $X$ and $P$ with non-vanishing commutator $\left[\hat{X}, \hat{P} \right] = i$. 
In an illustrative picture, the amplifier performs a measurement on the input 
state and subsequently prepares an amplified version in the amplification process. 
The measurement on the non-commuting observables results in the 3\,dB penalty, 
as e.g. characteristic in heterodyne detection \cite{Mueller16}.

Clearly, the same transformation also applies in the multi-stage amplifier 
to the state after the first amplifier, the only difference being the 
reduced gain of $\sqrt{G}$.
At the second amplifier stage, it is crucial to distinguish between 
the classical and the quantum mechanical components of the input variance. 
The bosonic Heisenberg uncertainty component experiences the 
identical transformation as in the first amplifier
$1 \cdot \mathrm{Var}_{\mathrm{coh}} \mapsto \left( 1 + 2(G-1) \right)\,\mathrm{Var}_{\mathrm{coh}}$, 
while the classical excess noise from the previous amplification 
stage does not suffer from the quantum penalty and hence just scales 
with the \textit{classical} gain factor $\sqrt{G}$: 
$\left( 2(\sqrt{G}-1) \right)\,\mathrm{Var}_{\mathrm{coh}} \mapsto \sqrt{G}\left( 2(\sqrt{G}-1) \right)\,\mathrm{Var}_{\mathrm{coh}} = 2G - 2\sqrt{G}$. 
Adding up the individual contributions yields the characteristic factor 
of any quantum-limited linear amplification $\left(2G-1\right)$ 
proving that the SNR in multi-stage amplifiers is indeed 
equivalent to just a single amplifier. 
As a consequence, consecutive high gain amplifications 
still yields only a 3\,dB SNR reduction compared to the input state 
Note, however, that larger degradation of the SNR occurs 
if the signal is subject to losses in between the amplifier stages.

\end{document}